# THE SUB-KEV ATOM REFLECTING ANALYZER (SARA) EXPERIMENT ABOARD CHANDRAYAAN-1 MISSION: INSTRUMENT AND OBSERVATIONS

ANIL BHARDWAJ*, MARTIN WIESER[†], M. B. DHANYA*, STAS BARABASH[†],
FUTAANA YOSHIFUMI[†], MATS HOLMSTRÖM[†], R. SRIDHARAN*, PETER
WURZ[#], AUDREY SCHAUFELBERGER[#], and ASAMURA KAZUSHI[¶]

*Space Physics Laboratory, Vikram Sarabhai Space Centre, Trivandrum, 695022,
Kerala, India
[†]Swedish Institute of Space Physics, Box 812, 98128, Kiruna, Sweden
[#]Physikalisches Institut, University of Bern, Sidlerstrasse 5, CH-3012, Bern, Switzerland
[¶]Institute of Space and Astronautical Science, 3-1-1 Yoshinodai, Sagamihara, Japan

SARA experiment aboard the first Indian lunar mission Chandrayaan-1 had the objective to explore the solar wind–lunar interaction using energetic neutral atoms (ENA) from the lunar surface as diagnostic tool. SARA consisted of an ENA imaging mass analyzer CENA (Chandrayaan-1 Energetic Neutral Analyzer) and an ion mass analyser SWIM (Solar Wind Monitor), along with a digital processing unit (DPU) which commands and controls the sensors and provides the interface to the spacecraft. Both sensors have provided excellent observational data. CENA has observed ENAs from the lunar surface and found that ~20% of the incident solar wind ions get backscattered as ENAs from the lunar surface. This is contrary to the previous assumptions of almost complete absorption of solar wind by the lunar surface. The observation is relevant for other airless bodies in the solar system.

## 1. Introduction

The solar wind, which is a continuous flow of plasma from the Sun, and consisting of ~96% $H^+$, ~4% $He^{++}$ and <0.1% heavier ions[1,2], interacts with obstacles in its flow such as planetary bodies or moons. For the Moon, the absence of a global magnetic field and the presence of a surface-bound exosphere makes the solar wind to interact directly with the lunar surface. The impacting solar wind ions, on interaction with the lunar regolith, get mostly neutralized. Particles can get absorbed on the surface, thermalized or directly backscattered to space[3,4,5]. As a part of this interaction, atoms can be ejected from the surface (a process known as sputtering) that have sufficient kinetic energy[6,7] to be called energetic neutral atoms (ENAs). Observation of backscattered or sputtered ENAs from the lunar surface along with the monitoring of impacting solar wind ions provides understanding of the lunar–solar wind-surface interaction[8].

Launch of Chandrayaan-1 on 22 October 2008 and its insertion into the lunar orbit on 8 November 2008 marked the beginning of a new era in the planetary exploration program of India. Chandrayaan-1 carried 11 experiments onboard – one among them was the Sub-keV Atom Reflecting Analyzer (SARA)[6,9], which used a novel technique of ENA imaging to investigate the lunar–solar wind interaction from a 100 km polar orbit. Since the ENAs released from the lunar surface can have energy greater than the escape energy of Moon, once released, they travel along straight lines and can be observed at 100 km altitude. Around magnetic anomalies, mini-magnetospheres may form[10]. In the presence of a mini-magnetosphere on the lunar surface, the direct entry of solar wind to that region will be obstructed, resulting in reduced ENA flux from that region[7,11]. Hence, magnetic anomalies present on the lunar surface can also be investigated with the SARA.

## 2. Instrumentation

The SARA experiment has two sensors, Chandrayaan-1 Energetic Neutrals Analyser (CENA) for measuring ENAs, and Solar Wind Monitor (SWIM) for simultaneous measurement of plasma in the lunar environment. The two sensors are connected to the Digital Processing Unit (DPU) which commands and controls the sensors. The block diagram of SARA, which also shows the flight models of CENA, SWIM and DPU, is shown in Figure 1. The total weight of the SARA is 4.4 kg and consumes a power of 17.1 W.

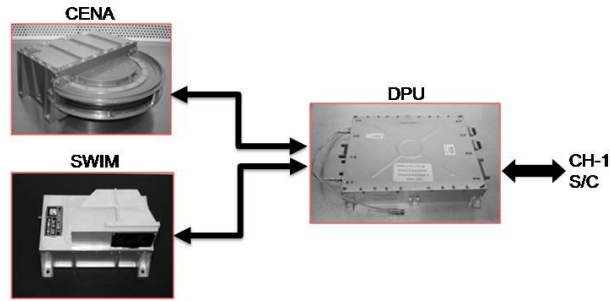

Figure 1. Block diagram of the SARA. The sensors CENA and SWIM, and the DPU are marked in the figure. CH-1 S/C refers to Chandrayaan-1 spacecraft.

CENA measures ENAs in the energy range 0.01 – 3.3 keV. Its principle is based on the conversion of neutral atoms to positive ions via surface-interaction technique. Its main functional blocks are an ion deflection system, a conversion surface, an electrostatic wave system for energy analysis, and a time-of-flight (TOF) section. The energy analysis combined with the velocity measurement in the TOF section allows mass determination[12]. The cross-sectional view of CENA is shown in Figure 2.

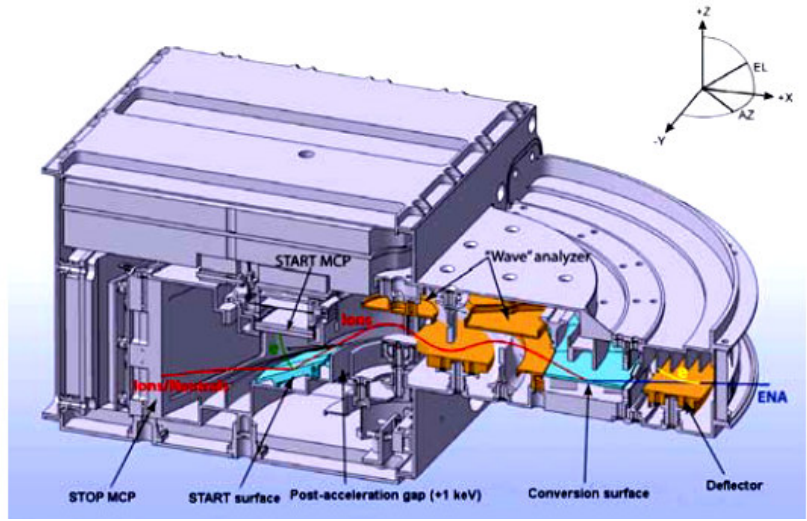

Figure 2. Cross-sectional 3D view of CENA with typical particle trajectory shown as solid line (taken from Ref. 9).

CENA has an electrostatic deflector near the entrance to deflect away charged particles entering the system. The deflector consists of two electrodes with one biased to +5 kV and the other at ground, which almost completely deflect away ions of energy <15 keV. Neutral particles, unaffected by the deflector, are incident at grazing incidence on a conversion surface that converts the neutral particles to positive ions. The conversion surface is a highly polished silicon wafer coated with MgO to provide high ionization efficiency[13]. The ions then pass through a wave-type electrostatic energy analyzer. The analyzer uses four different variable potentials applied to a system of electrodes to achieve energy analysis of the converted ions. The wave system also forms an optical labyrinth, efficiently blocking UV photons. Furthermore, the electrodes of the wave system are serrated and blackened to increase the UV suppression. This concept of the wave analyzer is similar to the one used in the MTOF sensor of the CELIAS instrument[14] on SOHO, which provides a photon rejection factor of $2\times10^{-8}$ (Refs. 6, 9). Before entering the TOF section, particles are post accelerated by a potential of 2.4 kV. In the TOF section, the converted ions fall on a start surface at 15° incidence and produce secondary electrons which propagate towards start MCP to produce a start pulse. The START surface is a highly polished mono-crystalline tungsten plate that provides specular particle reflection with low angular scattering. The surface material is also selected for high secondary electron yield and stable surface properties. The ions that are reflected from the start surface move mostly in form of neutral atoms toward the STOP MCP to produce the stop pulse. The difference in the timing of start and stop pulse gives the TOF. Since the distance the particle traverses between start surface and stop MCP is known from the instrument geometry, the velocity of the particle is calculated. As the energy of the ions from the energy analyzer is known with the information on post acceleration, the mass of the particle is deduced. The START MCP uses two sets of position sensitive anodes for the accurate determination of time-of-flight path length as well as to know the arrival direction of particles[6,9,12,15]. Characteristics of CENA are summarized in Table 1.

Table 1. Characteristics of CENA and SWIM

| Parameter | CENA | SWIM |
|---|---|---|
| Particle to measure | Neutrals | Ions |
| Energy range | 10 eV – 3.3 keV | 10 eV/q –15 keV/q |
| Energy resolution | 50 % | 7 % |
| Mass range (amu) | 1 – 56 | 1 – 40 |
| Mass resolution | H, O, Na/Mg/ Si/Al-group, K/Ca-group, Fe-group | $H^+$, $He^{++}$, $He^+$, $O^{++}$, $O^+$, > 20 amu |
| Full field-of-view | 15° × 160 ° | 9° × 180° |
| Angular resolution (FWHM) | 9°×25° (E >50 eV) | 4.5° × 22.5° |
| G-factor/sector without efficiency | $10^{-2}$ cm² sr eV/eV (at 3.3 keV) 0.2 cm² sr eV/eV (at 25 eV) | ~5 × $10^{-4}$ cm² sr eV/eV (0.5–3 keV) for $H^+$ |
| Efficiency (%) | 0.01 – 1 | 0.1 – 5 |
| Sensor mass | 1977 g | 452 g |

SWIM is an ion-mass analyzer; which is used for the observation of ions in the lunar vicinity in the energy range 0.01 – 15 keV/q. SWIM consists of an electrostatic deflector to scan the arrival direction of incident ions, an electrostatic energy analyzer and a time-of-flight section[16]. A schematic view of SWIM is shown in Figure 3.

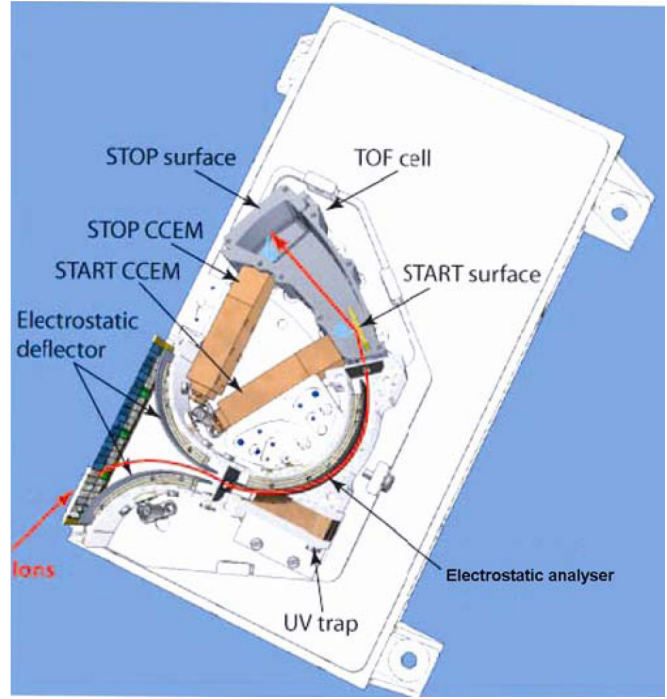

Figure 3.  3D view of SWIM with typical particle trajectory (taken from Ref. 9).

The electrostatic deflector consists of two electrodes. By suitably varying the voltages applied on the two plates, sweeping of the arrival direction of ions is done thereby achieving a total azimuthal coverage of 180°. Following the deflector, particles enter a cylindrical electrostatic analyzer with a deflection angle of 127°. Varying the voltages applied on one of its electrodes provides energy analysis of the ions. Additionally, at the entrance of the energy analyzer, a UV-trap reduces sensitivity to UV photons. After exiting the energy analyzer, the ions are post accelerated by 400 V and enter the TOF section. The ions fall on start surface at grazing incidence and produce secondary electrons which are detected by start CCEM (Ceramic Channel Electron Multipliers) to produce start pulse. The ions gets reflected from the start surface mainly as neutral atoms and fall on the STOP surface producing secondary electrons which are collected by the STOP CCEM to produce stop pulse. The difference in timing of start and stop pulse gives the TOF from which the mass per charge of the ions are calculated in a similar way as described for CENA above. As in CENA, the start surface is made of mono-crystalline polished tungsten. The STOP surface, made of graphite coated by MgO, is optimized for high secondary electron yield and high UV absorption. Characteristics of SWIM are summarized in Table 1.

CENA and SWIM are electrically interfaced with the spacecraft through the DPU (cf. Figure 1). The DPU commands and controls the sensors. It powers the sensors, sets the sensor modes and telemetry modes, acquires data from the sensors, does the time-integration and binning, formats the data, and transfers the data to telemetry. The DPU is built around ADSP21060 DSP processor running at 32 MHz clock frequency and having 4 Mb on-chip memory for operational software. Data is transferred from DPU to spacecraft through MIL-STD-1553 bus. Along with the science data, DC-DC converter monitoring, current monitoring for the sensor units and other sensor health check data of the scientific instruments are digitized and transmitted to spacecraft telemetry through 1553 bus[6,9].

SARA is mounted on the top deck of the 3-axis stabilized Chandrayaan-1 spacecraft such that CENA is nadir viewing. SWIM is mounted at nearly 90° to CENA. The full field of view (FOV) of CENA is 160° × 15° and that of SWIM is 180° × 9°. CENA has 7 and SWIM has 16 viewing directions. In this configuration, due to the 180° field of view of SWIM, a few viewing directions of SWIM will be looking at the lunar surface. The mounting geometry of SARA and the field of view of the sensors along with the orbital motion of Chandrayaan-1 are as shown in Figures 4(a) and 4(b).

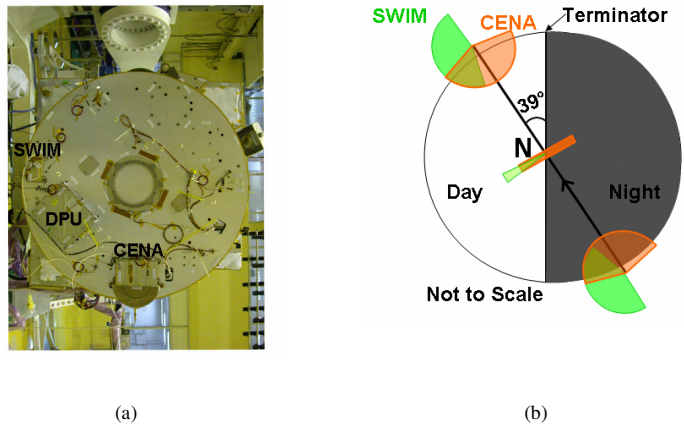

<div align="center">(a)           (b)</div>

Figure 4. (a) Mounting of CENA, SWIM and DPU on the top deck of the Chandrayaan-1 along with the FOV. (b) Field of view of CENA and SWIM at the poles and equator along with the orbit of Chandrayaan-1 for 6 February 2009. Day-night terminator is marked in the figure. CENA is nadir (Moon-facing) viewing and SWIM views at 90° to CENA.

## 3. First Observations and Results

Because high voltages were used in SARA, sufficient outgassing time was required before the first switch on. SARA was first switched on two months after the launch and commissioning was completed by the end of January 2009. Regular operations of SARA began on 5 February 2009.

An example of the ion data is shown in Figure 5: Energy spectra of ions as measured by SWIM during its first continuous observation on 25 January 2009 is shown. At this time, the Moon was upstream of the Earth's bowshock. SWIM measurements show that the solar wind has an energy of ~500 eV/q on this day. The peak corresponding to that of He$^{++}$ is seen around 1000 eV/q. The reflected solar wind ions show a broader energy spectrum which also peaks around 500 eV/q. Recently, Kaguya MAP/PACE has also observed reflected ions from lunar surface[17].

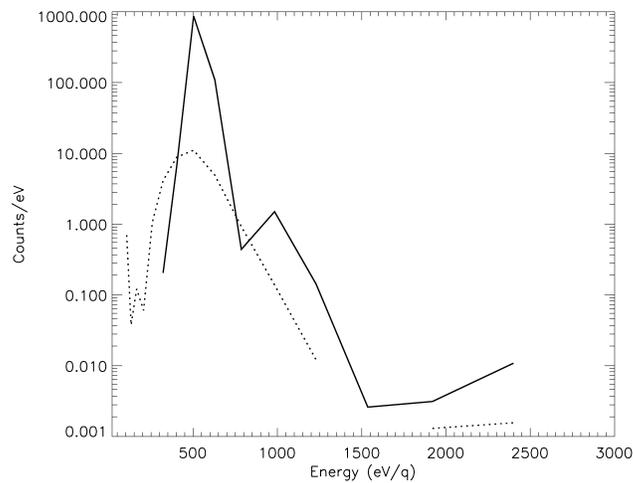

Figure 5. The energy spectra of solar wind ions observed by SWIM on 25 January 2009. The solid line represents the spectra of the impacting solar wind ions and the dotted line represents the spectra of the ions reflected from the lunar surface. The plot shows integrated counts for 36 minutes centered at 14:12 UT.

Another example of SARA data is shown in Figure 6, where energy spectra of impinging solar wind protons from SWIM observation are compared with energy spectra of reflected energetic neutral hydrogen atoms from CENA observation on 6 February 2009[18]. Data from three different orbits are shown. The reflected neutral flux varies accordingly with the intensity of the impinging solar wind ions.

The spectra of reflected hydrogen ENAs are broader and have a distinct upper energy of ~250 eV, after which the flux falls off rapidly. The solar wind protons have peak energy around 500 eV. The observed energy of ENAs indicate that solar wind protons lose significant amount of energy before getting backscattered as neutral hydrogen atoms. The fraction of reflected ENAs with respect to the flux of incident solar wind protons is found to be around 20%[18].

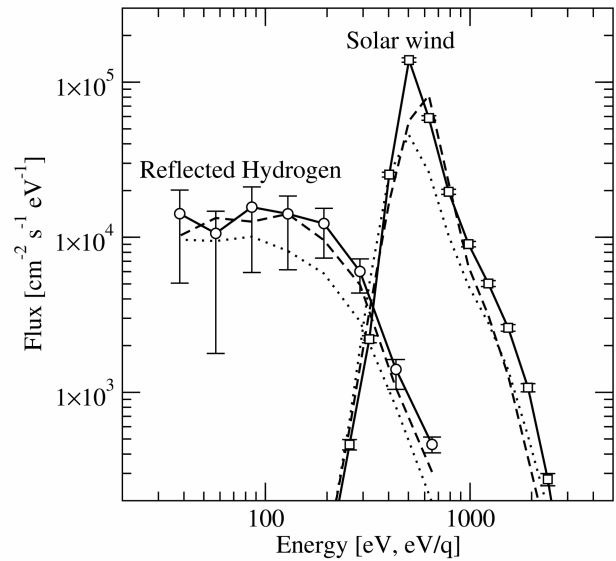

Figure 6. Flux of reflected hydrogen ENAs observed by CENA (circles) and solar wind protons observed by SWIM (square) for 6 February 2009 [taken from ref. 18]. The data is for 3 orbits which are indicated by the 3 lines each for backscattered hydrogen ENAs and solar wind protons. Solid lines represent the data for the orbit with dayside equator crossing at 05:22 UTC, dashed lines represent the orbit with dayside equator crossing at 07:20 UTC and the dotted lines represent the orbit with dayside equator crossing at 09:18 UTC. Error bars shown represent knowledge of the instrument geometric factor (1σ).

## 4. Discussion

CENA observations show that Moon is a strong source of ENAs. About 20% of the incident solar wind flux is found to be backscattered as ENAs from the lunar regolith. Most of these hydrogen ENAs have energy less than half of the impacting solar wind protons. Thus, SARA observations are in contrast with the earlier suggestions that only ~1% of incident solar wind ions are backscattered from lunar surface as neutral atoms[4]. The solar wind ions which are absorbed or implanted in the lunar regolith can get released to the lunar exosphere due to processes such as diffusion, solar wind sputtering and micrometeorite impact vaporization[19]. The CENA observation of high hydrogen reflection rate from lunar surface calls for a re-look at hydrogen inventory of the lunar regolith. The significant energy loss of the observed ENAs compared to impinging solar wind is due to elastic and inelastic processes during surface-plasma interactions, and possibly other processes such as retarding of solar wind ions by a positive surface potential. High hydrogen reflection is expected to occur on other regolith covered atmosphere-less bodies in the solar system, such as Mercury, asteroids, or Phobos. Recently, IBEX[8] also observed backscattered hydrogen ENAs from the Moon albeit from a much larger distance and found an ENA albedo of ~10 %. The IBEX results are consistent with the findings of the SARA.

The high hydrogen reflection may reduce the amount of solar wind hydrogen available for trapping in permanently shadowed areas near the poles[20,18]. This may have implications for one of the proposed mechanisms for the formation of water and OH in lunar regolith by the implantation of solar wind hydrogen[21]; the presence of water on lunar surface has recently been unambiguously evidenced by Moon Mineralogy Mapper experiment on Chandrayaan-1[22], and confirmed by observations made by Cassini VIMS[23] and Deep Impact's HRI-IR Spectrometer[24].

In addition to the measurement of direct solar wind, SWIM has observed solar wind ions backscattered from the lunar surface[25]. The reflected ions can be picked up by the solar wind convection electric field and get accelerated and may return back to the lunar surface again or can be lost. These ions may significantly contribute to the changes in the global plasma environment around the Moon[26].

### Acknowledgments

The efforts at Space Physics Laboratory, Vikram Sarabhai Space Centre, were supported by Indian Space Research Organization (ISRO), while the efforts at the Swedish Institute of Space Physics and University of Bern were supported in part by European Space Agency (ESA).